\documentclass[aps,preprint,superscriptaddress]{revtex4}
\usepackage{bm,graphicx,graphics,amsmath,amssymb,bm,epsfig,color}

\begin{document}
\bibliographystyle{apsrev}
\preprint{To appear in J. Appl. Phys. (manuscript CE-09)}

\title{Mechanical detection of FMR spectrum in a normally magnetized YIG disk}

\author{V. Charbois} 
\affiliation{Service de Physique de l'Etat Condens{\'e}, CEA Orme des
  Merisiers, F-91191 Gif-Sur-Yvette}

\author{V. V. Naletov} 
\affiliation{Service de Physique de l'Etat Condens{\'e}, CEA Orme des
  Merisiers, F-91191 Gif-Sur-Yvette}
\affiliation{Physics Department, Kazan State University, Kazan 420008 Russia}

\author{J. Ben Youssef} \affiliation{Laboratoire de Magn{\'e}tisme de
  Bretagne, CNRS/UPRESA 6135, 6 Av. Le Gorgeu, F-29285 Brest}

\author{O. Klein}
\thanks{Corresponding author} 
\email{oklein@cea.fr}
\affiliation{Service de Physique de l'Etat Condens{\'e}, CEA Orme des
  Merisiers, F-91191 Gif-Sur-Yvette}


\begin{abstract}
  The ferromagnetic resonance spectrum of a normally magnetized YIG disk,
  with thickness of 4.75$\mu$m and radius of 80$\mu$m, is measured at room
  temperature both by magnetic resonance force microscopy and by standard
  detection of the microwave susceptibility. The comparison indicates that
  MRFM represents one of the most potent means of obtaining the
  \emph{complete} FMR spectra of micron-size samples. In the weak coupling
  regime, the measured data can be quantitatively understood within the
  framework of the Damon and Eshbach model.
\end{abstract}

\pacs{PACS numbers: {07.79.Pk}{Magnetic force microscopes} \and
  {07.55.-w}{Magnetic components, instruments and techniques} \and
  {76.50.+g}{Ferromagnetic, antiferromagnetic, and ferrimagnetic
    resonances} }

\maketitle


There has been recently a renewed effort, both theoretical
\cite{albuquerque:01} and experimental \cite{hoffmann:91}, to describe the
spin dynamics of patterned magnetic thin films in anticipation of new fast
magnetic devices. The time scale of interest is the nanosecond because it
corresponds to the precession frequency of the magnetization about the
internal field direction. The classical method of exciting spin waves is
ferromagnetic resonance (FMR). These experiments study the splitting of the
ground state through the observation at microwave frequencies of continuous
wave absorption spectrum \cite{wigen:84}. But standard cavity methods
usually require sample dimensions that are much larger than the typical
lateral size of actual micro-fabricated device.

For micron-size samples \cite{zhang:97}, a new sensitive technique, called
ferro-Magnetic Resonance Force Microscopy (fMRFM) \cite{zhang:96}, has been
developed and its principle is to mechanically detect the net change of
dipole moment induced by FMR resonances. It uses a permanent micro-magnet
placed in the stray field of the sample, which couples to the longitudinal
magnetization (component along the static field, $M_z$) through the dipolar
interaction \cite{suh:98}. The force and torque acting on the probe
produces a measurable elastic bending of a cantilever onto which the magnet
is affixed.

The sensitivity of the mechanical detection will be illustrated with a
comparison of MRFM data and standard measurements of the microwave
susceptibility. In this paper, the experiments will be restricted to the
weak coupling regime, where a large separation ($h=100\mu$m) is set between
the probe and the sample.


Yttrium Iron Garnet (YIG) is the marvel material of FMR because of its
remarkably low loss in magnetic propagation \cite{damon:65}. An important
simplification of the MRFM spectroscopic signal can be obtained by using
disk samples, since a probe magnet placed on the disk axis will preserve
the axial symmetry. Cylindrical geometries have been extensively studied in
the sixties on millimeter-size rods and disks
\cite{dillon:60,eshbach:63,yukawa:74}. The measured resonances are ascribed
to magnetostatic waves propagating radially across the sample. Their
separation is determined by the cylinder aspect ratio \cite{dillon:60}.
They are labeled by $(n,m)$, the number of nodes respectively in the
diametrical and circumferential directions.  Magneto-exchange modes have
been found to be negligible for film thickness above 5$\mu$m \cite{wigen:84}.

\begin{figure}

\includegraphics*[scale=0.35, draft=false,clip=true]{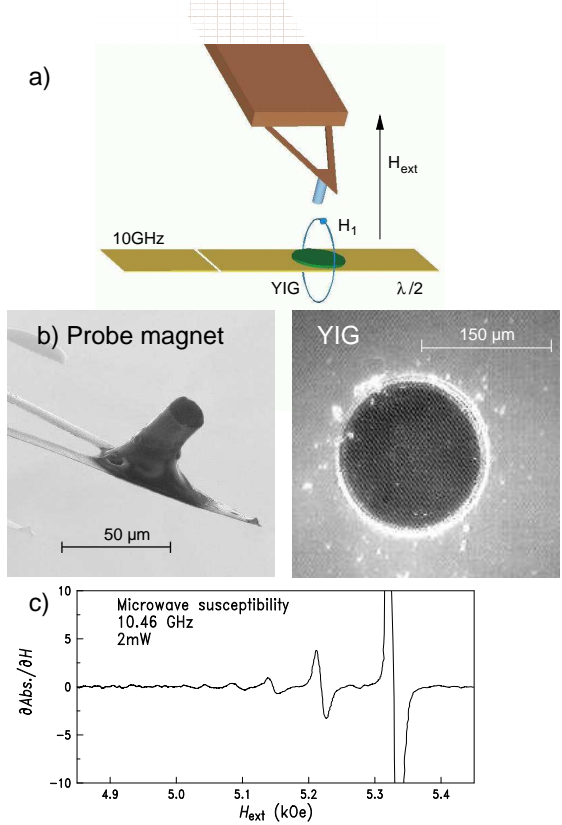}
\caption{
  a) Schematic (not to scale) of the setup geometry used for measuring the
  changes of the $z$ component of the magnetization as a function of the
  homogeneous static magnetic field, $H_{\text{ext}}$, parallel to the
  normal of a YIG disk. The disk is set on a half-wavelength strip-line
  resonator. $\Delta M_z$ is sensed by a magnetic bar glued on a cantilever and
  positioned $h=100\mu$m above the sample and on the disk axis b) Microscopy
  images of both the cylindrical probe magnet and the YIG disk. c)
  Imaginary part of the microwave susceptibility of the disk.}
\label{fig1}
\end{figure}

Our sample is micro-fabricated (ion milled) from a single crystal film
(thickness $S=4.75 \mu$m) oriented along the [111] direction (easy axis) into
a disk of radius $R\approx80\mu$m as shown in Fig.\ref{fig1}b. The dimensions are
large enough so that standard FMR experiments can be carried out on the
sample. Fig.\ref{fig1}c shows the microwave susceptibility of the disk as a
function of the dc magnetic field applied along the disk axis ($Oz$). The
transverse absorption spectrum ($\propto M_t$) is measured at 10.46GHz by a wide
band spectrometer using a standard crystal diode detector. The incident
power is 2mW which is about the maximum excitation possible while keeping
nonlinear effects reasonable \cite{seagle:85}. Four magnetostatic modes are
resolved.  They correspond to the longest wavelength spin-wave modes
allowed.

The same disk has then been used to test our mechanical FMR detection. The
setup is schematically represented in Fig.\ref{fig1}a. The MRFM fits
between the poles of an iron core electromagnet which applies a static
field $H_{\text{ext}}$ along the disk axis.  A proton NMR gaussmeter is
used for the calibration of $H_{\text{ext}}$. The microwave field $H_1$ is
generated by a 10.46GHz synthesizer and fed into an impedance matched
strip-line resonator of length $5$mm and width $0.5$mm. The YIG disk is
placed at the center of this half-wavelength resonator, with the 0.19mm
thick GGG substrate intercalated between the YIG and the strip-line. The
microwave field $H_1$ can be considered \emph{homogeneous} (within 4\%) over
the volume of the sample.  The sample temperature is fixed at $T=285$K and
the saturation magnetization is $4 \pi M_s(T) = 1815$G. A magnetic bar, 18
$\mu$m in diameter and 40 $\mu$m in length, \cite{acher:99} is glued (see
Fig.\ref{fig1}b) at the extremity of a soft cantilever (spring constant \(
k=0.01 \)N/m). SQUID measurements indicate that the room temperature
saturation magnetization of our bar is around $5\times10^2$emu/cm$^3$.  The bar
is lifted 110$\mu$m above the YIG in zero field.  A large separation is
purposely chosen, so that the coupling between the probe magnet and the
sample is weak (the magnetic field gradient produced on the sample is less
than $0.16\text{G}/\mu$m). This situation is obviously not optimal for
imaging purposes, but provides a meaningful comparison with the data
measured without the bar. The dipolar field acting on the bar produces an
{\AA}-scale displacement measured by a laser beam deflection on a photodiode.
When $H_{\text{ext}}$ reaches the 0.5 tesla field range, the cantilever
bends by 10$\mu$m towards the sample surface and the effective spring
constant stiffens to \( k=0.2 \)N/m.  The MRFM signal is proportional to
the changes of the longitudinal magnetization $\Delta M_z$ and thus it increases
linearly with microwave power ($\propto H_1^2$) below saturation (the transverse
component $M_t$ being proportional to $H_1$).

\begin{figure}
\includegraphics*[height=8cm,draft=false,clip=true,angle=90]{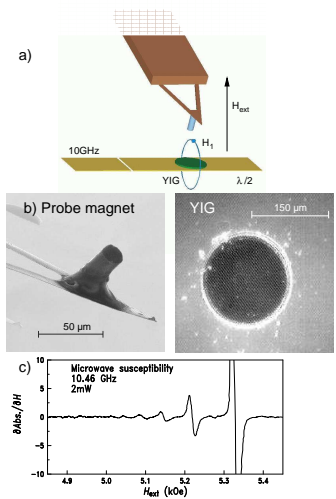}
\caption{Mechanically detected FMR spectrum of the
  normally magnetized YIG disk. The signal is proportional to the changes
  of the longitudinal component of the magnetization, $\Delta M_z$. The absence
  of even absorption peaks, $(2n,0)$, is a signature that the probe is
  placed precisely on the symmetry axis of the disk.}
\label{fig2}
\end{figure}

Fig.\ref{fig2} shows the field dependence of the mechanical signal when the
bar is placed on the symmetry axis of the disk and the amplitude of $H_1$
is fully modulated at the resonance frequency of the cantilever
$f_c\approx2.8$kHz. The microwave peak power is increased gradually during the
sweep, from 25$\mu$W for the longest wavelength modes up to 2.5mW for
$H_{\text{ext}} < 4.7$kOe. The normalized result is shown on a
\emph{logarithmic} scale. A series of 50 absorption peak is resolved by our
instrument thereby demonstrating the excellent sensitivity of mechanical
detection. The linewidth of the peaks is of the order of 1.5G, a typical
value for YIG disks \cite{eshbach:63}. In FMR, the linewidth is much
smaller than $\gamma \Delta H_i$, the field distribution inside the sample, because
of the propagating character of the spin waves throughout the sample. $\Delta
H_i$ in our disk is set by the dipolar field ($\approx 2 \pi M_s$) and the
additional broadening introduced by the probe magnet is comparatively
negligible (less than $6.2$G). These findings are in sharp contrast to
results obtained with paramagnetic resonance imaging where the excitation
is localized to the sheet satisfying the resonance condition which leads to
inhomogeneously broaden linewidth.

The theory of magnetostatic modes in thin films has been established
by Damon and Eshbach in 1961 \cite{damon:61} and modified by Damon and
van de Vaart \cite{damon:65}. The transverse mode dispersion relation
of forward volume waves can be expressed as:
\begin{equation}
k_t = \frac{2}{S} \frac{1}{\sqrt{p}} \tan^{-1}  \left(
    \frac{1}{\sqrt{p}} \right) \label{disp}
\end{equation}
with $p$ a parameter that equals to $p_0= \left\{ B_i H_i - \omega^2/\gamma^2\right\}
/\left\{ \omega^2/\gamma^2 - H_i^2\right\} $ in the absence of exchange ($\omega$ is the
frequency of excitation and $\gamma$ is the gyromagnetic ratio). $p$ depends on
the inhomogeneous internal fields, $H_i = H_{\text{ext}} + H_a - 4\pi M_s
n_{zz}$ and $B_i = H_i + 4\pi M_s$, where $n_{zz}$ is the depolarization
factor and $H_a=58$G the magneto-crystalline anisotropy field along the
[111] direction. An analytical expression for $n_{zz}(r,z)$ exists in the
case of uniformly magnetized cylinders \cite{joseph:65}. This expression,
however, neglects the rotation of the magnetization close to the edge, a
correction of the order $M_s/H_{\text{ext}} $. From Eq.\ref{disp}, we infer
that magnetostatic modes are propagating in the region where $p$ is
positive and are otherwise evanescent waves. One window of special interest
(Fig.\ref{fig2}a) is when $ 4.577\text{kOe} < H_{\text{ext}} < 5.375$kOe
which corresponds to $H_i < \omega/\gamma < \sqrt{H_i B_i}$ at $r=0$. There, the
propagating region is the central part of the disk. Excited spin waves
experience a force due to the internal field gradient and are accelerated
radially towards the center. The wave reflects at the circle $r=r_1$,
defined by $p=\infty$ ($\omega/\gamma = H_i$). Standing waves occur when the
phase shift over a period $4 \int_0^{r_1} k(r) dr$ is equal to $n \times 2 \pi$ (the
integral form incorporates the spatial nonuniformity). In this picture, the
internal field distribution affects the phase delay, {\it i.e.}  the locus
of the resonance, but not the linewidth. In Fig.\ref{fig2}b, we observe
another broad absorption regime for $H_{\text{ext}}<4.577$kOe which has
been reported by Eshbach\cite{eshbach:63} and corresponds to
magneto-elastic modes excited by spin waves localized at the edge.

\begin{figure}
\input{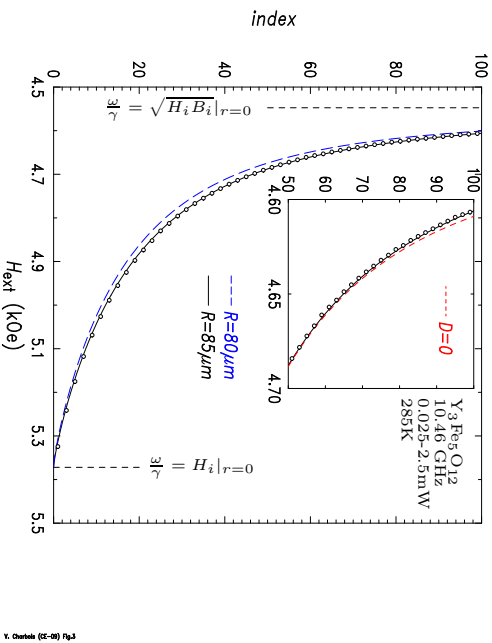}
\caption{Mode number $n$ as a function of the external
  field ($n$ is the number of standing waves in the diametrical direction).
  The open circles are the field position of each absorption peak measured
  in Fig.\ref{fig2}. The solid line is the theoretical predictions for a
  uniformly magnetized disk of radius $85\mu$m. The long dashed line is the
  same calculation for $R=80\mu$m (the physical dimension of the sample).
  The short dashed line in the inset shows the behavior when exchange
  effects are omitted ($D=0$).}
\label{fig3}
\end{figure}

We have used Eq.\ref{disp} to calculate $n$ as a function of
$H_{\text{ext}}$ for our sample. The resonance condition is calculated on
the median plan of the disk ($z=S/2$) and the result is shown as a long
dashed line on Fig.\ref{fig3}. There is {\it no} fitting parameter or
relative corrections. Our theoretical model, however, assumes a uniform
magnetization in the disk and thus underestimates the resonating field of
each mode (because it underestimates $n_{zz}$ \cite{joseph:65}).
Magnetization non-uniformities at the periphery of the disk introduce a
diminution of the radial decay of $H_i$ near the disk center. In our
picture, this is equivalent to an increase of the disk radius. We have
adjusted the radius $R=85\mu$m to fit the data at $n=30$ (the optimal fitting
range for the magnetostatic model).  For large $k$ ($n>50$), Eq.\ref{disp}
needs to be modified to include exchange effects. In our case, this effect
can be calculated as a first order correction of the characteristic
equation \cite{damon:65}:
\begin{equation}
k_t = k_t(p_0) - \frac{D}{\gamma \hbar} \frac{B_i +H_i (1 +2 p_0)}{B_i
    H_i - (\omega/\gamma)^2} \frac{k_t^2(p_0)}{S}  \left\{  \frac{S}{2}k_t(p_0) + \frac{1}{1+p_0} \right\} 
\end{equation}
with $D=0.93\times10^{-28}$erg.cm$^2$ the exchange parameter. The solid line
includes exchange effects and the dashed line in the inset shows the
behavior if $D$ is omitted. We have also calculated the alteration of the
spectrum due to the presence of the magnetic bar. For $h=100\mu$m, the
correction is small (less than 0.1\%), but the changes become more important
if the magnet is brought closer to the surface. The variation of the
spectrum with $h$ will be published elsewhere along with the details of the
calculation. For the sake of completeness, the solid line displayed is the
full result including the presence of the magnetic probe. The MRFM data
agree {\it quantitatively} with the model over the full range
\cite{damon:61}.

We are greatly indebted to C. Fermon, H. LeGall, O. Acher and A.L.
Adenot-Engeluin for their help and support in this work

\pagebreak

\pagebreak

\pagebreak

\end{document}